\documentclass[prl,showpacs,amsmath,amssymb,preprint]{revtex4-1}

\usepackage{graphicx}

\begin{document}

\title{A plasmon enhanced attosecond extreme ultraviolet source}

\author{Mattia Lupetti$^a$, Matthias F. Kling$^{b,c}$, and Armin Scrinzi$^a$}
\email{mattia.lupetti@physik.uni-muenchen.de}
\email{armin.scrinzi@lmu.de}
\affiliation{
$^a$Ludwig Maximilians Universit\"at, Munich, Germany\\
$^b$Max-Planck-Institut f\"{u}r Quantenoptik, Garching, Germany\\
$^c$J.R. Macdonald Laboratory, Kansas State University, Manhattan, USA}


\date{\today}

\begin{abstract}
  A compact high repetition rate attosecond light source based on a
  standard laser oscillator combined with plasmonic enhancement is presented. 
  At repetition rates of tens of MHz, we predict focusable pulses with durations of $\lesssim300$ attoseconds,
  and collimation angles   $\lesssim5^\circ$. Attosecond pulse parameters are robust 
  with respect variations of driver pulse focus and duration.   
\end{abstract}

\maketitle

High harmonic sources with controlled attosecond  ($1\,as=10^{-18}s$) 
time structure allow the observation of electronic dynamics on the natural 
time-scale of valence electrons.  A range of recently developed techniques for
ultrafast spectroscopy rely on such sources, for example, attosecond streaking of
photo-emission from atoms \cite{drescher:auger,remetter06_streaking}, and solid
surfaces \cite{cavalieri07:surface}, as well as attosecond transient absorption 
spectroscopy 
\cite{goulielmakis10:absorption}.

The attosecond pulse sources used in present experiments are all based on
high harmonic emission from gases. An extremely non-linear conversion
process --- ingeniously described by the classical three-step
re-collision model \cite{corkum93:simple-man,kulander93:simple-man} 
--- imprints the time-structure of the
driving laser pulse onto the harmonic radiation. Non-linearity
sharpens the original sine-like driver
electric field to produce bursts of high frequency radiation which may be as
short as 67 attoseconds \cite{zhao12:67attoseconds}.  
The bursts are time-delayed relative to the maxima of the driver field by 
$\sim 0.2$ driver optical cycles, i.e. emission occurs
near the {\em nodes} of the driver field. The time-locking
of the attosecond pulse to the driver is exploited for precision time-delay
experiments with time resolutions down to  $\sim 10\,as$.

Other than pulse duration, key parameters of the sources
are photon energy, pulse separation, peak and average intensity,
and for certain applications \cite{stockman07_attoplasmonics}, focal spot size.  
Maximal photon energies of high gas harmonics depend on
driver pulse intensity as $\sim I_p+3.2\,U_p$, where $I_p$ is the 
ionization potential of the gas and $U_p$ is the driver
ponderomotive potential.  This dependency is a direct consequence of
the re-collision mechanism for their generation.  For photon-energies
$\gtrsim 40\,eV$ with pulses at the Ti:sapphire wavelength of 800$\,nm$ one needs
intensities $\gtrsim 10^{14}W/cm^2$.  At driver pulse durations
$\sim20\,fs$ a short ``train'' of attosecond pulses is generated, i.e.\ a
sequence of a few pulses separated from each other by half the driver
optical period, as each half-cycle near the peak intensity creates its own
burst.
While trains of attosecond pulses are comparatively easy to generate,
using them for time-resolving processes that extend over more than 
half an optical period ($\sim 1.3 fs$) is difficult and requires  
deconvolution of the overlapping signals from neighboring pulses.

Isolated single attosecond pulses are generated by using extremely
short driver pulses where only one optical cycle is
strong enough to generate high harmonics at the relevant photon
energies \cite{baltuska03:nature}. Alternatively to short pulses, one can also manipulate 
polarization on the time scale of a single optical cycle in order to
suppress high harmonic generation for all but one field peak \cite{sola06:polarization_gating}. With this
mechanism, time-separation is given by the separation of subsequent driver
pulses rather than by the optical half-cycle.
These techniques are experimentally complex.  While laser
oscillators delivering pulses as short as 4 fs FWHM at 80~MHz repetition
rate are available \cite{rausch08:high_rep_rate}, amplification to $\gtrsim10^{14}\,W/cm^2$
brings the repetition rate down into the kilohertz regime.

In a recent paper \cite{park11:plasmon} plasmonic excitation in a cone-shaped silver funnel was proposed
to replace the  laser amplification stages and directly use
the high repetition rate oscillator for locally generating the 
intensities needed for high harmonics. Significant field
enhancement was theoretically predicted and extreme ultraviolet
(XUV) radiation was detected experimentally. A
follow-up theoretical study systematically investigated the dependence of the plasmonic field
on funnel geometry and drive pulse parameters \cite{choi12:plasmon_attosecond}. The fields
where found to produce attosecond time structure in the response of an isolated atom.

In the present work we investigate, whether the harmonic radiation is emitted from the funnel in 
the form of a usable beam and whether the time-structure is 
maintained in the macroscopic response. We find that emission from the narrow end of the funnel is
severely diffracted rendering it very hard to use in experiments. In contrast, in the reverse
direction from the large opening of the funnel a well-collimated beam with single attosecond
pulse structure is emitted. This ``plasmon enhanced attosecond XUV source'' (PEAX) is compared
to standard attosecond gas harmonic sources in terms of beam quality and yield.
Our analysis is based on solving the three-dimensional Maxwell equations for propagation of
driver, plasmon, and harmonic fields, and solving the atomic time-dependent
Schr\"odinger equation for the microscopic high harmonic response.

\begin{figure}
  \includegraphics[width=\columnwidth]{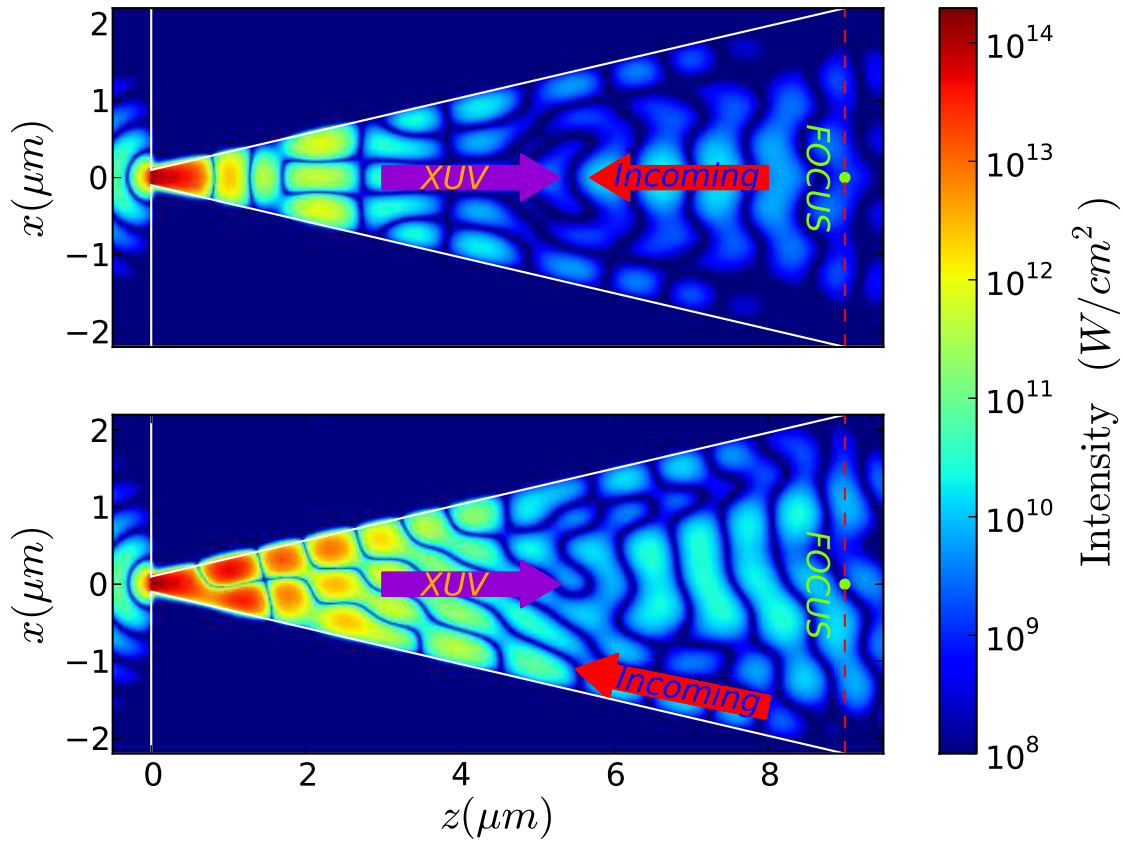}
  \caption{(color online) Intensity distribution in the silver cone  
    in the $xz$-plane at peak plasmon field of $2\times10^{14}W/cm^2$.
    Normal (upper panel) and oblique incidence (lower panel). 
    Polarization is in $y$-direction perpendicular to the plane.
    Oblique incidence angle is 7 degrees in the $zx$-plane parallel 
    to the inner cone surface. 
    In both cases, the harmonic beam is emitted in the direction
    of the cone axis.
  \label{fig:topology}}
\end{figure}

\begin{figure}[h]
  \includegraphics[width=\columnwidth]{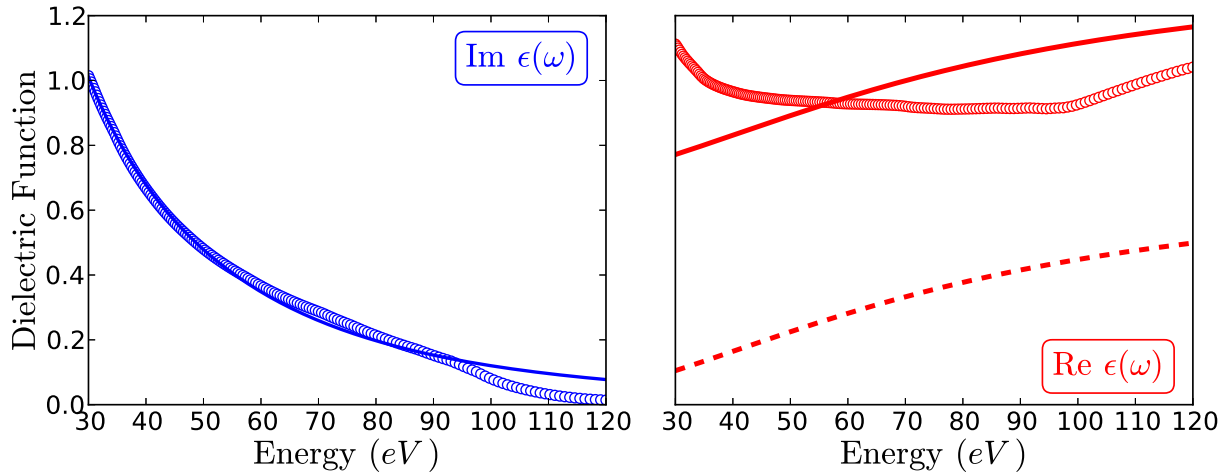}
  \caption{(color online) Dielectric functions used for the simulation of high
    harmonic propagation. Imaginary parts of the experimental values (left, blue circles) 
    are well approximated by a Lorentzian shape (blue line),
    The fit (right, red line) of the experimental real values (circles) is comparatively poor.
    The deliberately bad fit (dashed line) was used for checking robustness of the simulation (see text).
  \label{fig:dielectric-response}}
\end{figure}

Similar as in reference \cite{park11:plasmon} we chose for our simulations a $9\mu m$
long silver cone with elliptic cross-section and opening angles of 14 and 3.5 degrees
along the major ($x$-) and minor ($y$-)axis of the ellipse, respectively. We assume a 
5 fs FWHM Gaussian driver pulse at wavelength $\lambda_0=800\,nm$, beam waist $w_0=2.5\mu m$,
and focused intensity $I_0\approx4\times10^{11}W/cm^2$. The focal spot is placed at the larger 
opening of the cone (see Figure~\ref{fig:topology}). Variations of the 
focus position by $\pm1\mu m$ cause intensity changes of less than 5\% in peak plasmon intensity.
For solving Maxwell's equations we used the finite difference time-domain open source code MEEP 
\cite{MEEP}. For algorithmic reasons, dielectric functions in MEEP must have
Lorentzian shape, which poorly reproduces the known dielectric response of silver at
photon energies around $60\,eV$. Care was taken to accurately fit the imaginary part 
of the response (Fig.~\ref{fig:dielectric-response}). We found that the real part of the 
dielectric response has little influence on the results: with the two vastly 
different responses shown in Fig.~\ref{fig:dielectric-response} intensity and time-structure
of the final harmonic signal change by less than 15\%. 

As observed in Ref.~\cite{park11:plasmon}, the elliptic cross-section of the cone 
improves the plasmonic enhancement of the evanescent wave. With
strong eccentricity of $\epsilon=0.25$  we find
an enhancement of $\sim 500$  over the peak intensity of the 
incident driver field. 
With $\epsilon=0.5$ used in Ref.~\cite{park11:plasmon} the enhancement reduced by about a factor 3.
A similar dependence on ellipticity was reported in \cite{choi12:plasmon_attosecond},
where pulse durations between 4 and 10 fs were investigated. Note that peak field
is reached at the surface and therefor surface roughness will introduce modifications 
of the exact maxima. However, harmonics are produced in volumes of the scale of the evanescent wave,
which are less subjected to such sub-wavelength modifications.  
Optimizing the cone geometry allows for even weaker driver pulses. Note however, 
that ultimately the field inside the cone is limited by damage to the silver surface. 
Fortunately, in \cite{park11:plasmon} it was found that silver can support much higher fields than
expected, likely due to the extreme shortness of the pulses \cite{plech05_ablation}.

For high harmonic propagation, the driver field was sampled with
grid points separated by $2.5\,nm$,
well below the characteristic wavelength of 800 nm
of the driver plasmon in the cone and below the relevant harmonic wavelengths
of $\sim 27\,nm$. The atomic responses were obtained by
solving the time-dependent Schr\"odinger equation by the irECS method
\cite{scrinzi10:irecs} using a single-electron model with the ionization potential of Argon.
The harmonic yield drops rapidly with driver field intensity: 
from $2\times10^{14}W/cm^2$ to $1\times10^{14}W/cm^2$ the yield near the cutoff photon energy 
of $\sim 60\,eV$ drops by $\sim 3$ orders of magnitude.
We therefore restricted the calculation of the
responses to the volume $x\times y\times z\approx240\times60\times500\,nm^3$, 
where driver intensity exceeded $10^{14}W/cm^2$.

Coupling of the driver into the cone and harmonic emission can be maintained
over a range of incidence angles. Comparing to driver incidence on axis,
plasmonic enhancement drops by about a factor 4 
at incidence angle 7 degrees around the polarization ($y$-) axis, 
parallel to the inner cone surface at the major axis.
The smaller enhancement can be compensated by an increase in driver intensity.  

At short wavelength, numerical propagation through the complete cone in three
dimensions leads to an inappropriately large computational problem.
However, the influence of the cone on the propagation of short wavelength 
radiation can be expected to be small. We tested this assumption
in a wedge-shaped silver structure, where symmetry allows reduction to a two-dimensional
problem. 
We found that guiding in the silver cone improves collimation of 
the beam compared to propagation in vacuum, but leaves the time structure largely unaffected. 
Numerical convergence of the propagation in vacuum was verified  by direct numerical 
summation of the Li\'enard-Wiechert potentials of the individual atomic dipoles.
Propagation from the end of the cone to large distances 
is given by Kirchhoff's diffraction integral.

\begin{figure}[h]
  \includegraphics[width=\columnwidth]{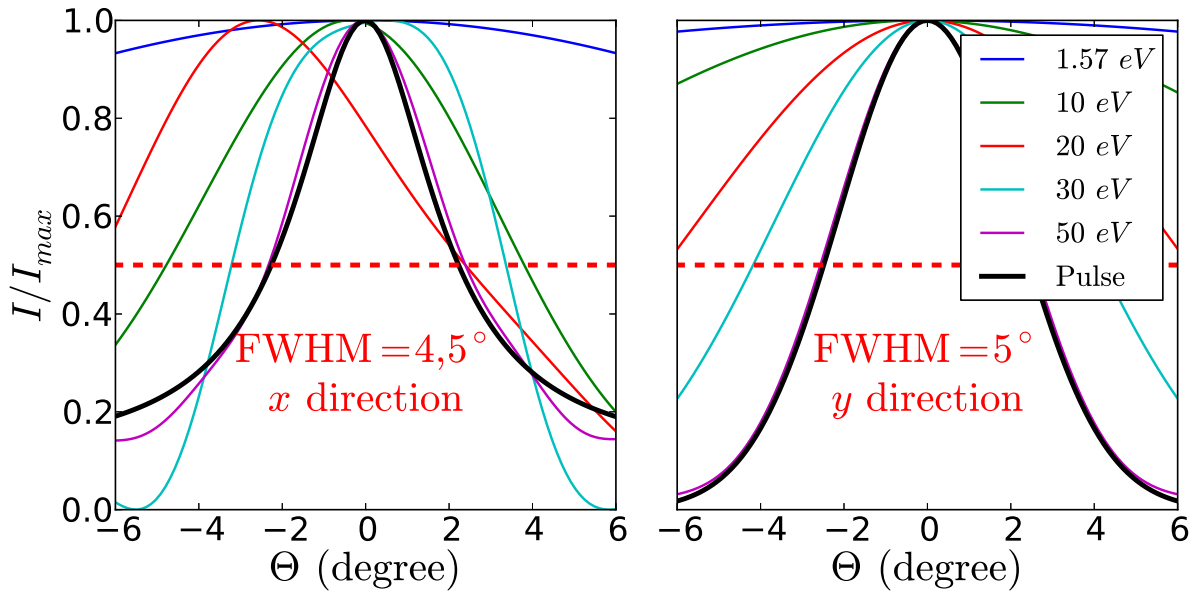}
  \caption{(color online) Far field angular distribution at a range of photon energies. The asymmetry in the left panel 
($x$-direction) is due to the oblique incidence. The right panel shows polarization- ($y$-) direction.
Profiles are taken at 1 mm distance from the cone. The thick black lines show divergence for a 
pulse composed of all harmonics above $45\,eV$.
  \label{fig:spatial-sep}}
\end{figure}
We compute the far field in the three-dimensional geometry
from the Kirchhoff integral with the MEEP solution at $z=2.5\mu m$.
Figure~\ref{fig:spatial-sep} shows the angular distribution of harmonic emission
out of the wide side of the cone for a range of harmonic frequencies at
oblique driver incidence. 
The incident intensity was adjusted such as to 
obtain a peak field in the cone of $2\times10^{14}W/cm^2$.
The beam divergence decreases with increasing photon energy, where the
divergence angle is defined by decrease to half the peak intensity.
For on-axis incidence, similar results are obtained, when the incident intensity 
is adjusted to give the same peak plasmon
intensity. The effect of the driver incidence angle on the emitted beam is small,
as the plasmon peak field is dominated by the same single plasmon mode, irrespective
of the exact incidence angle. 

With its extension $\lesssim 500\,nm$ the volume over which harmonics are generated
remains below the driver wavelength. Over that distance, free electron
dispersion does not affect phase matching of the harmonic with the dominant plasmon 
mode. Similarly, atomic dispersion is expected to remain small, and, where needed,
may be controlled by choosing a target gas suitable for a given harmonic wavelength.
Geometrically induced phase shifts are automatically included in the simulation.
With negligible phase slip between driver and harmonic across the generation volume,  
harmonic intensities grow quadratically with the gas density.

\begin{table}
\caption{\label{tab:cpas}
Harmonic beam characteristics for oblique incidence PEAX and a standard harmonic source
using a Gaussian beam (see text for parameters). 
Yields and photon flux are integrated over the beam divergence angles.
}
\begin{tabular}{| c | c | c |}
  \hline
  & PEAX & Gauss \\
  \hline
  Photon energy $\omega_{\gamma}$ &  $45\, eV$ & $45\, eV$ \\
  \hline
  Pulse duration $\Delta t$ & $300\, as $ & $250\, as$  \\
  \hline
  Rep Rate & $80\, MHz$ & $ 3\, kHz$ \\ 
  \hline
  Yield per pulse& $6.7 \cdot 10^{-3}$ $\gamma$/pulse & $ 3\cdot 10^{4}$ $\gamma$/pulse\\
  \hline
  Photon flux & $ 5.4 \cdot 10^5\, s^{-1}$ & $9 \cdot 10^7\, s^{-1}$\\
  \hline
  Beam divergence & $5^\circ$ & $1^\circ$\\
 \hline
  Active volume & $\sim4\times10^{-3}(\mu m)^3$ & $\sim16 (\mu m)^3$\\
 \hline
\end{tabular}
\end{table}

The first column of Table~\ref{tab:cpas}
shows the parameters of the harmonic pulses obtained at gas pressure of $0.3$ bar
(density $7.8\cdot 10^{18} cm^{-3}$), a value typically used for standard gas harmonics.
Due to the rapid decay of the spectral intensity with harmonic energy, the central frequency
nearly coincides with the lower cutoff frequency of the harmonics.
The PEAX harmonics above $45\,eV$ form an isolated attosecond pulse.
Pulse contrast is satisfactory with 85\% of the energy in the central peak.

The harmonic pulse is emitted with a perfectly spherical wave front. At a distance 
of $1\,mm$ and over its divergence angle of $5^\circ$, the deviation from spherical 
shape remains below 5\% of the central pulse wavelength. This can be ascribed to
the very small and well-defined ``focal spot'', i.e. the single plasmonic mode
where the high harmonics are generated.
The clean wave front allows for focusing without compromising the time-structure.
The attosecond pulse is emitted on the cone axis, while  
the reflected driver and lower harmonics are emitted into wider angles with 
a modulated intensity profile (cf.~Fig.~\ref{fig:spatial-sep}). This allows simple geometric separation of the 
incident driver pulse from the harmonic pulse. Remaining on-axis components 
of the reflected driver and low harmonics can be blocked by standard filters.

Emission through the narrow end of the cone is comparable in photon-flux, 
but beam divergence is $35^\circ$ in $y$-direction, making it hard to focus
for experimental use. 

The data given is for a 5~fs driver pulse. A shorter, 4~fs driver 
bears no further advantage: harmonic pulse duration and divergence remain 
unchanged, but emitted intensity is reduced by about 20\%.
 Similarly, normal incidence of the driver does not
improve harmonic pulse parameters, but rather leads to a reduction of intensity 
because of the smaller active volume (cf. Fig.~\ref{fig:topology}).

In the second column of Table~\ref{tab:cpas}, the pulse parameters for
a traditional harmonic source are listed. 
For the comparison a Gaussian beam with 4\,fs FWHM pulse duration was used.
We assumed a tight focus in a gas jet with peak intensity in focus
equal to the peak plasmonic intensity of $2\times10^{14}W/cm^2$ and a beam waist of $1\mu m$. 
At the Rayleigh length of $3.9\mu m$, the volume for producing high harmonics is
by three orders of magnitude larger than in PEAX. At our specific parameters,
PEAX shows slightly longer pulse duration and larger beam
divergence.  Note that PEAX beam divergence is
overestimated due to the omission of large part of the funnel in 
harmonic propagation. At photon energies $\sim 45\,eV$ and at
the given gas pressure, the photon yield per shot is almost 7 orders
of magnitude larger compared to PEAX.  This ratio corresponds to a
ratio of a bit more than 3 orders of magnitude in field strengths,
which is consistent with the ratio of the active volumes of the two
sources. Due to the low required intensity, driver pulses for PEAX can be directly
drawn from a laser oscillator at repetition rates near 100~MHz,
reducing the ratio of photon-fluxes between the two sources to $\lesssim500$.
With the very small active volume in PEAX, gas density can be
increased significantly before an optical thickness is reached where
phase matching problems arise and coherence of the source
deteriorates. Assuming tenfold pressure for the PEAX, photon flux
could be boosted by 2 orders of magnitude, basically closing the gap
to the traditional gas harmonic source. Another obvious extension of
PEAX would be to use an array of many instead of a single cone. Only
recently, geometries with multiple cones were realized experimentally
\cite{kim_private12}. In such an arrangement, interference between the 
emission from different cones could be utilized for phase-matching.

However, it must be said that volumes for coherent gas harmonic generation can be significantly
larger than what was assumed in our example. With proper placement of the beam focus relative to the gas
jet, phase matching can be maintained over about 1 mm propagation length at a beam cross section
of a few hundred $\mu m^2$, further increasing volumes by 3-5 orders of magnitude over our example.
This may turn out to be a fundamental limitation for intensities from any plasmon enhanced harmonic 
source: the natural length 
scale of the high intensity spots in such sources is the driver wavelength, while much large diameters
and phase matching length can be realized in gas harmonic sources. 
The motivation for using PEAX therefore rather 
arises from other advantages, such as its sub-micrometer collimated attosecond XUV beam, which may be used for 
spatio-time-resolved surface spectroscopy \cite{stockman07_attoplasmonics} and the high repetition rates that 
reduce space charge effects on the target. In principle, with the high repetition rate and possibly few-cycle 
pulses in the $45\,eV$ ($\sim 27\,nm$ wavelength) regime, frequency combs could be pushed to new parameter regimes.
As plasmonic fields may vary over very spatial ranges
non-dipole modifications of the atom-field interactions are introduced that may be exploited, e.g., for further 
reduction of pulse duration and for higher photon energies \cite{shaaran12:non-dipole}. 

Technically, the PEAX appears simple as no strong few-cycle
pulses are required. The mechanism is robust with respect to small variations in driver pulse focusing
and duration. The weak dependence on incidence angles allows geometrical separation
of the driver pulse from the attosecond pulse. Without the need for short pulse amplification 
and with its high repetition rate, PEAX is an attractive alternative source that may become 
accessible also outside specialized short pulse laboratories.

We acknowledge support by the DFG, by the excellence cluster "Munich Center for Advanced
Photonics (MAP)", and by the Austrian Science Foundation project ViCoM
(F41). M.~L. is a fellow of the IMPRS ``Advanced Photon
Science''.  M.F.K. acknowledges support by the U.S. Department of Energy 
under DE-SC0008146, the BMBF via PhoNa, and the DFG via Kl-1439/4 and
Kl-1439/5.  We are grateful for discussions with V.~Yakovlev.

\end{document}